\documentclass[runningheads]{llncs}

\usepackage{graphicx}
\usepackage[ruled,noend,linesnumbered]{algorithm2e}
\usepackage{amssymb,amsmath}
\usepackage{diagbox}
\usepackage{url}

\DeclareMathOperator*{\argmin}{arg\,min}
\DeclareMathOperator*{\argmax}{arg\,max}
\newcommand{\calF}{{\mathcal{F}}}
\newcommand{\calN}{{\mathcal{N}}}

\newcommand{\A}{A}
\newcommand{\pospo}{pos(p,p_o)}
\newcommand{\pos}[2]{{pos(#1,#2)}}
\newcommand{\ptm}{{p_t^M}}

\def\set#1{\{#1\}}

\begin{document}
\title{Strategic Voting in the Context of Negotiating Teams}

\author{Leora Schmerler \and
Noam Hazon
%
%
\institute{Department of Computer Science, Ariel University, Israel \\
\email{\{leoras,noamh\}@ariel.ac.il}}}

\maketitle

\begin{abstract}
A negotiating team is a group of two or more agents who join together as a single negotiating party because they share a common goal related to the negotiation. Since a negotiating team is composed of several stakeholders, represented as a single negotiating party, there is need for a voting rule for the team to reach decisions. In this paper, we investigate the problem of strategic voting in the context of negotiating teams. Specifically, we present a polynomial-time algorithm that finds a manipulation for a single voter when using a positional scoring rule. We show that the problem is still tractable when there is a coalition of manipulators that uses a $x$-approval rule. The coalitional manipulation problem becomes computationally hard when using Borda, but we provide a polynomial-time algorithm with the following guarantee: given a manipulable instance with $k$ manipulators, the algorithm finds a successful manipulation with at most one additional manipulator. Our results hold for both constructive and destructive manipulations.
\end{abstract}

\section{Introduction}

Voting is a common way to combine the preferences of several agents in order to reach a consensus. While being prevalent in human societies, it has also played a major role in multi-agent systems for applied tasks such as multi-agent planning~\cite{ephrati1993multi} or aggregating search results from the web~\cite{dwork2001rank}. In its essence, a voting process consists of several voters along with their ranking of the candidates, and a voting rule, which needs to decide on a winning candidate or on a winning ranking of the candidates.

Another common mechanism for reaching an agreement among several agents is a negotiation~\cite{fatima2014principles}. In a negotiation there is a dialogue between several agents in order to reach an agreement that is beneficial for all of them. 
Extensive work has been invested in developing negotiation protocols for many settings, but
bilateral negotiations, where there are only two negotiating parties, is the most common type of negotiations~\cite{brams2003negotiation}. 
Many works have focused on the case where each negotiating party represents a single agent. However, there are many cases in which a negotiating party represents more than one individual.

For example (motivated by S{\'{a}}nchez-Anguix et al.~\cite{sanchez2011analyzing}), consider an agricultural cooperative that negotiates with the government. 
Even though the members of the cooperative have a common goal, they may have different preferences regarding the prohibition of importing products, government supervision of prices, insect control, tax concessions, etc. 
As another example, consider the government of the United Kingdom that negotiates with the European Union (EU) regarding withdrawal from the EU (i.e., the Brexit). The members of the EU have similar interests and objectives, and thus they are considered a single party in the negotiation process. Nevertheless, the EU is composed of different countries, and they may have different preferences regarding sovereignty, migrants and welfare benefits, economic governance, competitiveness, etc. 
%
%
%
These situations are denoted by social scientists as negotiating teams,
in which \textit{a group of two or more interdependent persons join together as a single negotiating party because their similar interests and objectives relate to the negotiation}~\cite{brodt2001negotiating}.

Since a negotiating team comprises several stakeholders represented as a single negotiating party, there is need for a coordination mechanism, and a voting rule is a natural candidate.
%
Ideally, the voters report their true preferences so that the voting rule will be able to choose the most appropriate outcome. However, as shown by Gibbard~\cite{gibbard1973manipulation} and Satterthwaite~\cite{satterthwaite1975strategy}, 
every reasonable voting rule with at least $3$ candidates is prone to strategic voting. That is, voters might benefit from reporting rankings different from their true ones. Clearly, this problem of manipulation also exists in a negotiating team. For example, suppose that there is a EU council committee that negotiates with the UK on agricultural and fishery policies. 
The committee may decide that the UK will be excluded from the agricultural policy due to Brexit, or the UK will still be included. Similarly, the committee may decide that the fishery policy no longer applies to the UK or include the UK. Therefore, there are $4$ possible outcomes, denoted by $o_1,o_2,o_3$ and $o_4$. Now, suppose that Germany prefers $o_1$ over $o_2$, $o_2$ over $o_3$, and $o_3$ over $o_4$. We may assume that the preferences of the UK government are publicly known, and it is also possible that Germany, which currently holds the presidency of the EU council, is familiar with the preferences of the other EU council members. Since the negotiation protocol usually is also known, Germany might be able to reason that $o_3$ is the negotiation result, but if Germany will vote strategically and misreport its preferences then $o_2$ will be the negotiating result. To the best of our knowledge, the analysis of manipulation in the context of negotiating teams has not been investigated to date. 

In this paper, we investigate manipulation in the context of negotiating teams. We assume that there is a negotiation process between two parties. One of the parties is a negotiating team, and the team uses a voting rule to reach a decision regarding its negotiation strategy. Specifically, the negotiating team uses a positional scoring rule as a \emph{social welfare function (SWF)}, which outputs a complete preference order. This preference order represents the negotiating party, and is the input in the negotiation process. We thus assume that there is a negotiation protocol that can work with ordinal preferences. We use the \emph{Voting by Alternating Offers and Vetoes (VAOV)}  protocol \cite{anbarci1993noncooperative}, since it is intuitive, easy to understand, and the negotiation result is Pareto optimal. 
Moreover, Erlich et al.~\cite{erlich2018negotiation}  
have shown that we can identify the negotiation result of the VAOV protocol if both parties follow a sub-game perfect equilibrium with an intuitive procedure.

We analyze two types of manipulation, constructive and destructive. We begin by studying constructive manipulation by a single voter, where there is a single manipulator that would like to manipulate the election so that a preferred candidate will be the negotiation result. We show that 
placing the preferred candidate in the highest position in the manipulative vote is not always the optimal strategy, unlike in the traditional constructive manipulation of scoring rules, and we provide a polynomial-time algorithm to find a manipulation (or decide that such a manipulation does not exist).  
We then analyze the constructive coalitional manipulation problem, where several voters collude and coordinate their votes so that an agreed candidate will be the negotiation result. We show that this problem is still tractable for any $x$-approval rule, but it becomes computationally hard for Borda. However, we provide a polynomial-time algorithm for the coalitional manipulation of Borda with the following guarantee: given a manipulable instance with $k$ manipulators, the algorithm finds a successful manipulation with at most one additional manipulator.
Finally, we show that our hardness result and algorithms can be adapted for destructive manipulation problems, where the goal of the manipulation is to prevent a candidate from being the result of the negotiation.

The contribution of this work is twofold. First, it provides an analysis of a voting manipulation in the context of negotiating teams, a problem that has not been investigated to date. Our analysis also emphasizes the importance of analyzing voting rules within an actual context, because it leads to new insights and a deeper understanding of the voting rules. Second, our work concerns the manipulation of SWF, which has been scarcely investigated.

\section{Related Work}
The computational analysis of voting manipulation was initially performed by Bartholdi, Tovey, and Trick~\cite{bartholdi1989computational}, and Bartholdi and Orlin~\cite{bartholdi1991single}, who investigated constructive manipulation by a single voter. 
Following these pioneer works, many researchers have investigated the computational complexity of manipulation, and studied different types of manipulation with different voting rules in varied settings. We refer the reader to the survey provided by  
\cite{faliszewski2010ai}, and more recent survey by \cite{conitzer2016barriers}. All of the works that are surveyed in these papers analyze the manipulation of voting rules as \textit{social choice functions}, that is, the voting rules are used to output one winning candidate (or a set of tied winning candidates). In our work we investigate manipulation of a resolute SWF, i.e., it outputs a complete preference order of the candidates.


There are very few papers that investigate the manipulation of SWFs. This is possibly since the opportunities for manipulation are not well-defined without additional assumptions. That is, since the output of a SWF is an order, and voters do not report their preferences over all possible orderings, some assumptions have to be made on how the voters compare possible orders. Indeed, the first work that directly deals with the manipulation of SWF was by \cite{bossert1992strategy}, who assumed that a voter prefers one order over another if the former is closer to her own preferences than the latter according to the Kemeny distance, 
and mainly presented impossibility results.
Bossert and Sprumont~\cite{bossert2014strategy}
assumed that a voter prefers one order over another if the former is strictly between the latter and the voter's own preferences. Built on this definition their work studies three classes of SWF that are not prone to manipulation (i.e., strategy-proof).
Dogan and Lain{\'{e}}~\cite{dogan2016strategic} 
characterized the conditions to be imposed on SWFs so that if we extend the preferences of the voters to preferences over orders in specific ways the SWFs will not be prone to manipulation. Our work also investigates the manipulation of SWF, but we analyze the SWF in the specific context of a negotiation. Therefore, unlike all of the above works, the preferences of the manipulators are well-defined and no additional assumptions are needed.

Our work is also connected to committee elections or multi-winner elections, where manipulation of scoring rules has been considered \cite{meir2008complexity,obraztsova2013manipulation,bredereck2017coalitional}. However, in committee election we are given the size of the committee as an input.
%
In our setting the output of the voting rule (i.e., the ranking) essentially determines the point in which $RC$ terminates (see Section~\ref{sec:prem} for the definition of $RC$). Using the model of committee election in our setting we can say that the ranking determines the size of the committee. That is, each possible manipulation determines not only the position of each candidate but also the size of the committee.

The work that is closest to ours is the paper by S{\'{a}}nchez-Anguix et al.~\cite{sanchez2011analyzing}, which involves the use of voting rules for the decision process of a negotiating team, i.e., the same basic scenario that we consider. The paper presents several strategies they developed, which use some specific, tailored-made, voting rules, and experimentally analyzes them in different environments. Our work analyzes voting in the context of a negotiation from a theoretical perspective. We formally define the general problem, show polynomial-time algorithms for some cases, and provide hardness results and approximations for others.

Finally, we note that in our setting there is a SWF, which outputs an order over the candidates, and this order is used as an input for the negotiation process. In Section~\ref{sec:prem} we note that there is a connection between the sub-game perfect equilibrium of the negotiation and the Bucklin voting rule. Therefore, our setting is also related to a multi-stage voting.
Several variants of multi-stage voting have been considered \cite{conitzer2003universal,elkind2005hybrid,davies2012eliminating,narodytska2013manipulating}.
%
All of these works did not consider the case of SWF in the first round, as we do. More importantly, in all of these works the set of voters remains the same throughout the application of the voting rules. In our case the set of the voters in the first stage is different from the set in the second stage. In the first stage the voters are the agents in the negotiating team, and they use a scoring rule as a SWF. In the second stage there are only two voters, which are the negotiating parties, and they use an equivalent of Bucklin on their full preference orders.

\section{Preliminaries} \label{sec:prem}

We assume that there is a set of outcomes, $O$, $|O|=m$ and a set of voters $V=\set{1,...,n}$. Each voter $i$ is represented by her preference $p_i$, which is a total order over $O$. We write $o \succ_{p_i} o'$ to denote that outcome $o$ is preferred over outcome $o'$ according to $p_i$. The position of outcome $o$ in preference $p_i$, denoted by $\pos{o}{p_i}$, is the number of outcomes that $o$ is preferred over them in $p_i$. That is, the most preferred outcome is in position $m-1$ and the least preferred outcome is in position $0$
\footnote{Our definition of a candidate's position in a voter's ranking is the opposite of the commonly used, and we chose it to enhance the readability of the proofs: $\pos{o}{p_i} \geq \pos{o'}{p_j}$ is naturally translated to ``$o$ is ranked in $p_i$ higher than $o'$ is ranked in $p_j$''.}. We also refer to the outcomes of $O$ as candidates, and to the total orders over $O$ as votes. 

A preference profile is a vector $\vec{p} = (p_1,p_2,...,p_n)$.
In our setting we are interested in a \textit{resolute social welfare function}, which is a mapping of the set of all preference profiles to a single strict preference order.
A scoring vector for $m$ candidates is $\vec{s} = (s_{m-1},\ldots,s_0)$, where every $s_i$ is a real number, $s_{m-1} \geq \ldots \geq s_0 \geq 0$, and $s_{m-1} > s_0$. A scoring vector essentially defines a voting rule for $m$ candidates: each voter awards $s_i$ points to the candidate in position $i$.
Then, when using the rule as a SWF, the candidate with the highest aggregated score is 
placed in the top-most position, the candidate with the second highest score is placed in the second highest position, etc.
Since ties are possible, we assume that a lexicographical tie-breaking rule is used.
We study {\em positional scoring rules}, where each rule in this family applies an appropriate scoring vector for each number of candidates. That is, a scoring rule is represented by an efficiently compu   table function $f$ such that for each $m \in \mathbb{N}$, $f(m) = (s_{m-1}^m,\ldots,s_0^m)$ is a scoring vector for $m$ candidates.
Some of our results hold only for $x$-approval rules, in which  $f(m)=(1,\ldots,1,0,\ldots,0)$, where the number of $1$’s is $x$. Note that the well-known {\em Plurality} rule (where each voter awards one point to her favorite candidate) is $1$-approval and the {\em Veto} rule (where each voter awards one point to all the candidates, except for the least preferred one) is $(m$-$1)$-approval and they are thus both $x$-approval rules. We also analyze the Borda rule, where each voter awards the candidate a score that equals the candidate's position, i.e., $f(m)=(m-1,m-2,\ldots,1,0)$. In general, we denote the resulting social welfare function $\calF$.

In the negotiation process we assume there are two parties: $t$ is the negotiating team, which comprises a set of voters, and there is another party. The parties negotiate over the set of outcomes $O$, and their preferences are also total orders over $O$. However, since $t$ is a negotiating team that  comprises several stakeholders, the preference order of $t$, $p_t$, is determined by the social welfare function over the preference profile of the members of $t$, that is, $p_t = \calF(\vec{p})$. We denote by $p_o$ the preference order of the other party.

We assume that negotiating parties use the \textit{Voting by Alternating Offers and Vetoes (VAOV)} protocol~\cite{anbarci1993noncooperative}, which is a negotiation protocol that works with ordinal preferences. The protocol works as follows. Let $p_1$ be the party that initiates the negotiation and let $p_2$ be the other party. At round $1$, party $p_1$ offers an outcome $o \in O$ to $p_2$. If $p_2$ accepts, the negotiation terminates successfully with $o$ as the result of the negotiation. Otherwise, party $p_2$ offers an outcome $o' \in O \setminus \{o\}$.
If $p_1$ accepts, the negotiation terminates successfully with $o'$ as the result of the negotiation. Otherwise, $p_1$ offers an outcome
$o'' \in O \setminus \{o,o'\}$ to $p_2$, and so on. If no offer was accepted until round $m$ then the last available outcome is accepted in the last round as the result of the negotiation.
%
We further assume that the negotiating parties are rational and each party has full information on the other party's preferences. Therefore, the parties will follow a sub-game perfect equilibrium (SPE) during the negotiation. 
Anbarci~\cite{anbarci1993noncooperative} 
showed that if both parties follow an SPE the negotiation result will be unique. We can thus also call this outcome the SPE result. The SPE result depends on $p_t$, $p_o$, and on the identity of the party that initiates the negotiation, and we thus denote by $\calN_t(p_t,p_o)$ the SPE result if the negotiation team $t$ initiates the negotiation, and by 
$\calN_o(p_t,p_o)$ the SPE result if the other party initiates the negotiation.

In some negotiation settings there is a central authority that can force the parties to offer specific outcomes in a specific order. In this case it is common to use a bargaining rule, which is a function that assigns each negotiation instance a subset of the outcomes that is considered the result of the negotiation. One such bargaining rule is the \textit{Rational Compromise ($RC$)} bargaining rule~\cite{kibris2007bargaining}. Let $A^j_{(p_t)}=\{$the $j$ most preferred outcomes in $p_t \}$. $A^j_{(p_o)}$ is defined similarly for $p_o$. 
$RC$ is computed as follows:
\begin{enumerate}
	\item Let $j=1$
	\item If $|A^j_{(p_t)} \cap A^j_{(p_o)}| > 0$ then return $A^j_{(p_t)} \cap A^j_{(p_o)}$.
	\item Else, $j \leftarrow j+1$ and go to line $2$. 
\end{enumerate}
Note that the $RC$ bargaining rule is equivalent to Bucklin voting with two voters and no tie-breaking mechanism. 
An important finding of~\cite{erlich2018negotiation} shows that the negotiation result of the VAOV protocol if both parties follow an SPE (i.e., the SPE result) is always part of the set returned by the $RC$ rule. We use this connection between $RC$ and the VAOV negotiation protocol whenever we need to identify the SPE result.
Specifically, if $RC$ returns one outcome, this is also the SPE result. If $RC$ returns two outcomes then the SPE result depends on the number of outcomes and on the party that initiates the negotiation.
%
%

\section{Constructive Manipulation by a Single Voter} \label{sec:singleManipulator}

We begin by studying the problem of constructive manipulation by a single voter.
In this setting a manipulator $v'$ would like to manipulate the election so that a preferred candidate $p$ will be the SPE result. We assume that the decision of which party initiates the negotiation is not always known in advance. Therefore, we require that both $\calN_t$ and $\calN_o$ returns the preferred candidate.
The Constructive Manipulation in the context of Negotiations (C-MaNego) is defined as follows:
\begin{definition}[C-MaNego]
    We are given social welfare function $\calF$, a preference profile $\vec{p}$ of honest voters on the negotiating team $t$, the preference of the other party $p_o$, a specific manipulator $v'$, and a preferred candidate $p \in O$. We are asked whether a preference order $p_{v'}$ exists for the manipulator $v'$ such that $\calN_t(\calF(\vec{p} \cup p_{v'}),p_o) = \calN_o(\calF(\vec{p} \cup p_{v'}),p_o) = p$.
\end{definition}

We first observe that manipulation problems in the context of negotiations are inherently different from the traditional voting manipulation problems. First, in voting manipulation there is one set of voters in which their preferences are the inputs of the voting rule. The manipulator only needs to take these preferences into account when she decides on her manipulative vote. In our case there are two stages: in the first stage there is a set of voters and in the second stage there are two negotiating parties, and the manipulator needs to consider the preferences of all of these agents when she decides on her manipulative vote.
In addition, unlike constructive manipulation in many voting rules, placing the preferred candidate $p$ in the highest position in the manipulative vote is not always the optimal strategy, since constructive manipulation in our case requires sometimes also destructive actions. Indeed, the following example describes a scenario where there is no manipulation where $p$ is placed in the highest position. However, manipulation is possible if $p$ is placed in the second highest position, since this placement allows for a destructive action against another candidate.  
\begin{example}
Assume that $p_o$ is the following preference order: $
p_o = b \succ p \succ a \succ c.
$
There is one manipulator $v'$, and $\vec{p}$ comprises $4$  voters with the following preferences:
$p \succ c \succ a \succ b $, $p \succ b \succ a \succ c $, $b \succ p \succ a \succ c $, $b \succ a \succ c \succ p $.
Assume that we use the Borda rule, and thus the voters of $\vec{p}$ give the following scores: $b$ gets $8$ points, $p$ gets $8$ points, $a$ gets $5$ points and $c$ gets $3$ points. Since we assume that the tie-breaking rule is a lexicographical order, $p_t = b \succ p \succ a \succ c$. 
In order to find a successful manipulation $v'$ needs to make sure that $b$ will not be in the two highest positions in $\calF(\vec{p} \cup p_{v'})$. 
Now, if the manipulator places $p$ in the highest position then $p$ gets $11$ points. Then, placing the other candidates in every possible order results in $b$ in the second highest positions in $\calF(\vec{p} \cup p_{v'})$. Alternatively, if $v'$ votes as follows:
$
a \succ p \succ c \succ b,
$
then $p$ gets $10$ points, $a$ and $b$ get $8$ points, and $c$ gets $4$ points;  thus $\calF(\vec{p} \cup p_{v'}) = p \succ a \succ b \succ c$. Now the SPE result is $p$.
\end{example}

We now present a polynomial-time algorithm for C-MaNego with any scoring rule.
Let $p^a$ be the order that the algorithm finds (i.e., $p^a$ is a possible $p_{v'}$), and let $p_t^a = \calF(\vec{p} \cup p^a)$.
Note that during the algorithm we use $\calF(\vec{p} \cup p^a)$, where $p^a$ is not a complete preference order, i.e.,
$p^a$ comprises $m'$ candidates, $m'<m$, that are placed in specific positions.
In these situations we assume that all of the candidates that are not in $p^a$ get a score of $0$ from $p^a$. Given $i$, $1 \leq i \leq \lceil m/2 \rceil$, let $H^i$ be the set that contains $p$, and the $i-1$ most preferred outcomes in $p_t$ that do not belong to $\A^i_{(p_o)}$.

Our algorithm works as follows. 
It uses the connection between $RC$ and the negotiation protocol to identify the SPE result. Clearly, if the position of $p$ in $p_o$ is less than $\lceil m/2 \rceil$ then for any possible $p^a$ $RC$ does not return $p$. Therefore, there is no manipulation and the algorithm returns false (lines~\ref{line:1}-\ref{line:2}). Otherwise, we use the variable $i$ to indicate the iteration number in which $RC$ terminates. Thus, the algorithm iterates over the values of $i$ from $1$ to $\lceil m/2 \rceil$ (line~\ref{line:terminate_Rc}).
For a given $i$, the algorithm tries to ensure that no outcome from $\A^i_{(p_o)}$ will be placed in the $i$ highest positions in $p_t^a$. Consequently, the algorithm places the outcomes from $H^i$ in the highest positions and they receive the highest scores. Moreover, the outcomes are placed in a reverse order (with regards to their order in $p_t$) to ensure that even the least preferred outcome in $H^i$ will receive a score that is as high as possible (in order to be included in the highest positions in $p_t^a$). Then, the algorithm places the remaining outcomes, denoted $C$, so that they will not prevent $p$ from being the negotiation result (lines~\ref{line:define_C}-\ref{line:notH_end}). 
Specifically, the algorithm places the the outcomes of $C$ in the lowest positions in $p^a$ and the outcomes are placed in a reverse order, with regards to their order in $p_t$.
Then, if $p^a$ is a successful manipulation the algorithm returns it (line~\ref{line:return}). Otherwise, the algorithm proceeds to the next iteration.

%
%
\begin{algorithm}[hbpt]
\caption{Constructive manipulation by a single voter}
\label{alg:single_manip}
\SetAlgoLined
\LinesNumbered

    \If{$\pospo < \lceil m/2 \rceil$} 
    { \label{line:1}
        \textbf{return} false \label{line:2}
    }

    \For{$i = 1$ \text{ to } $\lceil m/2 \rceil$} 
    { \label{line:terminate_Rc}
        $p^a \leftarrow H^i$ in a reverse order of the positions in $p_t$\\
        $C$ $\leftarrow O \setminus H^i$ \label{line:define_C}\\
            \For{$j=1$ to $|O \setminus H^i|$} { \label{line:notH_begin}
                $c \leftarrow$ the most preferred outcome from $C$ under $p_t$ \\
                place $c$ in $p^a$ such that $\pos{c}{p^a} = j-1$ \\
                $j \leftarrow j+1$ \\
                remove $c$ from $C$
            \label{line:notH_end}
            }
        \If{$\calN_t(\calF(\vec{p} \cup p^a),p_o) = \calN_o(\calF(\vec{p} \cup p^a),p_o) = p$}
        {  \label{line:alg1_end_cond}
            \textbf{return} $p^a$
            \label{line:return}
        }
    }
    \textbf{return} false
\end{algorithm}
\begin{theorem}
\label{thm:alg_1_correct}
Algorithm~\ref{alg:single_manip} correctly decides the C-MaNego problem with any positional scoring rule in polynomial time.
\end{theorem}


\begin{proof}
Clearly, the algorithm runs in polynomial time since there are two loops, where each loop iterates at most $m$ times. In addition, if the algorithm successfully constructs a manipulation order, $p$ will be the negotiation result. We need to show that if an order that makes $p$ the negotiation result exists, then our algorithm will find such an order.
Assume that we have a manipulative vote, $p^m$, that makes $p$ the negotiation result, and let $p_t^m =  \calF(\vec{p} \cup p^m)$. Thus, $\calN_t(p_t^m,p_o) = \calN_o(p_t^m,p_o) = p$. In addition, given a set $H^i$ let $L^i = \set{\ell | \exists h \in H^i \text{ s.t. } h \prec_{p_t} \ell}$ and $R^i = \{o | o \in O, o \notin H^i \text{ and } o \notin L^i \}$.

We show that Algorithm~\ref{alg:single_manip} returns $p^a$ in line~\ref{line:return}, when $i$ equals the iteration in which $RC$ terminates given $p_t^m$ and $p_o$. 
There are two possible cases to consider:

    $\bullet$ \hspace{2pt} $\A_{(p^a)}^i=\A_{(p^m)}^i$ : 
    according to Algorithm~\ref{alg:single_manip}, $\A_{(p^a)}^i = H^i$, and since $\A_{(p^a)}^i=\A_{(p^m)}^i$, $\A_{(p^m)}^i = H^i$.
    By definition, $\forall r \in R^i$ and $\forall h \in H^i$, $r \prec_{p_t} h$ and $r \prec_{p^m}  h$. Since we use a scoring rule, $\forall r \in R^i$ and $\forall h \in H^i$, $r \prec_{p^m_t}  h$. Since $p^m$ is a successful manipulation and $RC$ terminates at iteration $i$, then $\forall \ell \in L^i$ where $\ell \in \A_{(p_o)}^i$, $\ell \notin \A_{(p_t^m)}^i$. For any other $\ell \in L^i$ we know that $p \prec_{p_t} \ell$ and for any $h \in H^i \setminus \set{p}$, $\ell \prec_{p_t} h$. Since $p^m$ is a successful manipulation and $RC$ terminates at iteration $i$, $p \in \A_{(p_t^m)}^i$. Overall, $\A_{(p_t^m)}^i = H^i$.

    We first assume that all the candidates that are not in $H^i$ get a score of $0$ from $p^a$, and we show that $\A_{(p_t^a)}^i = H^i$. For any $h \in H^i$, if $\pos{h}{p^a} \geq \pos{h}{p^m}$ then $\pos{h}{p_t^a} \geq \pos{h}{p_t^m}$. Otherwise, let $h \in H^i$ be a candidate such that $\pos{h}{p^a} < \pos{h}{p^m}$ and let $s = \pos{h}{p^a}$. There are $m-s-1$ candidates from $H^i$ above $h$ in $p^a$. According to the pigeonhole principle, at least one of them, denoted $h'$, that is placed in $p^m$ at position $s$ or lower. That is, $\pos{h'}{p^m} \leq \pos{h}{p^a}$. By the algorithm construction, all of the candidates that are ranked higher than $h$ in $p^a$ are ranked lower than $h$ in $p_t$. That is, $h' \prec_{p_t} h$. However, $h' \in \A_{(p_t^m)}^i$ and thus $h \in \A_{(p_t^a)}^i$. Overall, $\A_{(p_t^a)}^i = H^i$. 
%
    
    We now show that Algorithm~\ref{alg:single_manip}  (lines~\ref{line:notH_begin}-\ref{line:notH_end}) can assign scores to all the candidates in $O \setminus H^i$ such that $p^a$ is a successful manipulation. 
    For any $o \in O \setminus H^i$, if $\pos{o}{p^a} \leq \pos{o}{p^m}$ then $\pos{o}{p_t^a} \leq \pos{o}{p_t^m}$. Since $o \notin \A^i_{(p_t^m)}$ then $o \notin \A^i_{(p_t^a)}$. Otherwise, let $o \in O \setminus H^i$ be a candidate such that $\pos{o}{p^a} > \pos{o}{p^m}$ and let $s = \pos{o}{p^a}$. There are $s$ candidates from $O \setminus H^i$ below $o$ in $p^a$. 
    According to the pigeonhole principle, at least one of them, denoted $o'$, is placed in $p^m$ at position $s$ or higher. That is, $\pos{o'}{p^m} \geq \pos{o}{p^a}$. By the algorithm construction, all of the candidates $c \in O \setminus H^i$ that are ranked lower than $o$ in $p^a$ are ranked higher than $o$ in $p_t$. That is, $o \prec_{p_t} o'$. However, $o' \notin \A^i_{(p_t^m)}$ and thus $o \notin \A^i_{(p_t^a)}$. Overall, after placing the candidates from $O \setminus H^i$ in $p^a$, $\forall o \in O \setminus H^i$, $o \notin \A^i_{(p_t^a)}$. That is, $\A_{(p_t^a)}^i = H^i$, and thus $\calN_t(\calF(\vec{p} \cup p^a),p_o) = \calN_o(\calF(\vec{p} \cup p^a),p_o) = p$.

    $\bullet$ \hspace{2pt} $\A_{(p^a)}^i \neq \A_{(p^m)}^i$:
%
    let $p^{m\prime}$ be the manipulation $p^m$ with the following changes:  each $r \in \A_{(p_t^m)}^i \setminus H^i$ is replaced with a candidate $h_r \in H^i \setminus \A_{(p_t^m)}^i$. 
    That is, $\pos{r}{p^{m\prime}} = \pos{h_r}{p^m}$ and $\pos{h_r}{p^{m\prime}} = \pos{r}{p^m}$.
    Since $p^m$ is a successful manipulation, if $r \in \A_{(p_t^m)}^i \setminus H^i$ then $r \notin \A_{(p_o)}^i$. Thus, by the definition of $H^i$, $\forall r \in \A_{(p_t^m)}^i \setminus H^i$ and $\forall h \in H^i \setminus \A_{(p_t^m)}^i$, $\pos{r}{p_t} < \pos{h}{p_t}$. 
    Therefore, since each $r \in \A_{(p_t^m)}^i \setminus H^i$ is ranked in the highest $i$ positions in $p_t^m$, then $h_r$ is ranked in the highest $i$ positions in $p_t^{m\prime}$. 
    Similarly, since each $h_r$ is not ranked in the highest $i$ positions in $p_t^m$, then $r$ is not ranked in the highest $i$ positions in $p_t^{m\prime}$.
    That is, $h_r \in \A_{(p_t^{m\prime})}^i$ and $r \notin \A_{(p_t^{m\prime})}^i$, and thus, $H^i = \A_{(p_t^{m\prime})}^i$.
    %
    Let $p^{m\prime\prime}$ be the manipulation $p^{m\prime}$ with the following changes: each $r \in \A_{(p^{m\prime})}^i \setminus H^i$ is replaced with a candidate $h_r \in H^i \setminus \A_{(p^{m\prime})}^i$. 
    That is, $\A_{(p^{m\prime\prime})}^i = H^i$.
    Note that $c \notin \A_{(p_t^{m\prime})}^i$ for every $c \in O \setminus H^i$, and therefore $c \notin \A_{(p_t^{m\prime\prime})}^i$. Thus, $\A_{(p_t^{m\prime\prime})}^i = H^i$. That is, $p^{m\prime\prime}$ is a successful manipulation, and $\A_{(p^{m\prime\prime})}^i=\A_{(p^a)}^i$. This brings us back to the first case we already considered and showed that $p^a$ is a successful manipulation.
\qed \end{proof}

\section{Constructive Coalitional Manipulation}
We now consider the problem of constructive manipulation by a coalition of voters. That is, several manipulators, denoted by $M$, might decide to collude and coordinate their votes in such a way that an agreed candidate $p$ will be the SPE result.
The constructive coalitional manipulation problem is defined as follows:
\begin{definition}[CC-MaNego]
    Given a social welfare function $\calF$, a preference profile $\vec{p}$ of honest voters on the negotiating team $t$, the preference of the other party $p_o$, a number of manipulators $k$, and a preferred candidate $p \in O$,  we check whether a preference profile $\vec{p}_M$ for the manipulators exists such that $\calN_t(\calF(\vec{p} \cup \vec{p}_M),p_o) = \calN_o(\calF(\vec{p} \cup \vec{p}_M),p_o) = p$.
\end{definition}

We show that CC-MaNego can be decided in polynomial time for any $x$-approval rule using Algorithm~\ref{alg:coalitional_manip}, which works as follows.
\begin{algorithm}[hbpt]
\caption{Coalitional manipulation}
\label{alg:coalitional_manip}
\SetAlgoLined
\LinesNumbered

    \If{$\pospo < \lceil m/2 \rceil$} { \label{line:check_possibility}
        \textbf{return} false \label{line:not_possible}
    }
    \For{$i = 1$ to $\lceil m/2 \rceil$}{
        $\vec{p}_M \leftarrow []$ \\
        \For{$\ell=1$ to $|M|$}{ \label{alg:beginStage} 
            $p^a \leftarrow$ empty preference order \\
            $C \leftarrow H^i$ \\
            \For{$j=1$ to $|H^i|$} {
                $c \leftarrow$ the least preferred outcome from $C$ under $\calF(\vec{p} \cup \vec{p}_M)$ \\
                place $c$ in $p^a$ such that $\pos{c}{p^a} = m-j$ \\
                $j \leftarrow j+1$ \\
                remove $c$ from $C$
            }
            $C \leftarrow O \setminus H^i$ \\
            \For{$j=1$ to $|O \setminus H^i|$} {
                $c \leftarrow$ the most preferred outcome from $C$ under $\calF(\vec{p} \cup \vec{p}_M)$ \\
                place $c$ in $p^a$ such that $\pos{c}{p^a} = j-1$ \\
                $j \leftarrow j+1$ \\
                remove $c$ from $C$
            }
            add $p^a$ to $\vec{p}_M$
        }\label{alg:endStage}
        \If{$\calN_t(\calF(\vec{p} \cup \vec{p}_M),p_o) = \calN_o(\calF(\vec{p} \cup \vec{p}_M),p_o) = p$}{ \label{line:check_success}
            \textbf{return} $\vec{p}_M$ \label{line:success}
        }
    }
    \textbf{return} false
\end{algorithm}
Similarly to Algorithm~\ref{alg:single_manip}, the algorithm iterates over the possible values of $i$, where $i$ indicates the iteration number in which $RC$ terminates. For any given $i$, the algorithm iterates  over the number of manipulators and determines their votes (Lines \ref{alg:beginStage}-\ref{alg:endStage}). We refer to each of these iterations as a \emph{stage} of the algorithm. In each stage, a vote of one manipulator is determined, denoted by $p^a$.
We begin with an empty set of votes, $\vec{p}_M$. Then, the algorithm places the outcomes from $H^i$ in the highest positions in $p^a$. The outcomes are placed in a reverse order, with regards to their order in $\calF(\vec{p} \cup \vec{p}_M)$. Similarly, the algorithm places all the other outcomes in the lowest positions in $p^a$ and the outcomes are placed in a reverse order, with regards to their order in $\calF(\vec{p} \cup \vec{p}_M)$. Note that the set $H^i$ does not change throughout the algorithm's stages. However, the order of the outcomes in $H^i$ and $O \setminus H^i$ according to $\calF(\vec{p} \cup \vec{p}_M)$ may change when we update $\vec{p}_M$, which implies that the order in which we place the outcomes from $H^i$ and $O \setminus H^i$ in $p^a$ may differ from one vote to another.
%
\begin{theorem}
\label{thm:coalitional_k_approval}
Algorithm~\ref{alg:coalitional_manip} correctly decides the CC-MaNego problem with $x$-approval rule in polynomial time.
\end{theorem}

In order to prove Theorem~\ref{thm:coalitional_k_approval} we use the following definitions.
Recall that $k = |M|$. Let $s_\ell(c)$ be the score of candidate $c$ in $\calF(\vec{p} \cup \vec{p}_M)$ after stage $\ell$.
Given $i$, $1 \leq i \leq \lceil m/2 \rceil$, let $U^i_0 = \argmin_{h \in H^i} \pos{h}{p_t}$.
For each $s=1,2,...$, let $U^i_s \subseteq H^i$ be $U^i_s=U^i_{s-1} \cup \set{u: u$ was ranked above some $u' \in U^i_{s-1}$ in some stage $l$, $1 \leq l < k$, but $u$ was ranked below some $u' \in U^i_{s-1}$ in stage $l + 1}$. 
Now, let $U^i= \bigcup _{0 \leq s}U^i_s$.
The set $D^i$ is defined similarly. Specifically, let $D^i_0 = \argmax_{d \in O \setminus H^i}\pos{d}{p_t}$. For each $s=1,2,...$, let $D^i_s \subseteq O \setminus H^i$ be $D^i_s=D^i_{s-1} \cup \set{d: d$ was ranked below some $d' \in D^i_{s-1}$ in some stage $l$, $1 \leq l < k$, but $d$ was ranked above some $d' \in D^i_{s-1}$ in stage $l + 1}$. Now, let $D^i= \bigcup _{0 \leq s}D^i_s$. 
Note that $\forall s$ $U^i_s \neq U^i_{s-1}$ and $D^i_s \neq D^i_{s-1}$, and $s$ does not necessarily equal $k$.

We  begin  by  proving  some  Lemmas  that  are  necessary for the proof of Theorem \ref{thm:coalitional_k_approval}. 

\begin{lemma} \label{lemma:k_ranking}
Given $i$, $1 \leq i \leq \lceil m/2 \rceil$,
\begin{enumerate}
    \item The candidates in $U^i$ are placed in each stage $l$, $1 \leq l \leq k$ in the $|U^i|$ highest positions.
    \item The candidates in $D^i$ are placed in each stage $l$, $1 \leq l \leq k$ in the $|D^i|$ lowest positions.
\end{enumerate}
\end{lemma}

\begin{proof}
Assume by contradiction that there exists some $h \in H^i \setminus U^i$ that was placed in some stage in one of the first $|U^i|$ places, then there exists some $u \in U^i$ that was placed below $h$ at this stage. Let $s \geq 0$ such that $u \in U^i_s$. By definition, $h \in U^i_{s+1}$ and thus $h \in U^i$, which is a contradiction to the choice of $h$. The proof for the set $D^i$ is similar. 
\qed \end{proof}

\begin{lemma} \label{lemma:conditions_for_D=1_U=1}
Given $i$, $1 \leq i \leq \lceil m/2 \rceil$, if $\forall u \in U^i$ in each stage $j$, $u$
receives $1$ point, then $|U^i|=1$. Similarly, if $\forall d \in D^i$ in each stage $j$, $d$
receives $0$ points, then $|D^i|=1$. 
\end{lemma}

\begin{proof}
Let ${u_0} = U^i_0$. Assume to contradiction that there exists some $u \in U^i$ such that $u \neq u_0$. By definition of the set $U^i$, after some stage, $u$ was positioned lower than $u_0$. But $u$ and $u_0$ gained in each stage $1$ point, and therefore $u$ could not have bean positioned lower than $u_0$ in any stage. Contradiction to the existence of such a candidate $u$. The proof for the set $D^i$ is similar.
\qed \end{proof}

\begin{lemma} \label{lemma:k_score_difference_of_candidates}
Given $i$, $1 \leq i \leq \lceil m/2 \rceil$, for all $u_1, u_2 \in U^i$, $|s_k(u_1) - s_k(u_2)| \leq 1$. Similarly, for all $d_1, d_2 \in D^i$, $|s_k(d_1) - s_k(d_2)| \leq 1$.
\end{lemma}

\begin{proof}
Recall that $U^i_0 = \argmin_{h \in H^i} \pos{h}{p_t}$, and for each $s=1,2,...$, $U^i_s=U^i_{s-1} \cup \set{u: u$ was ranked above some $u' \in U^i_{s-1}$ in some stage $l$, $1 \leq l < k$, but $u$ was ranked below some $u' \in U^i_{s-1}$ in stage $l + 1}$. In addition,  $U^i= \bigcup _{0 \leq s}U^i_s$.
We prove by induction on the index $s$, that for all $u_1, u_2 \in U^i$, $|s_k(u_1) - s_k(u_2)| \leq 1$. In the base case, when $s=0$, $|U^i_0|=1$ and thus the inequality trivially holds. Assume that for $s-1$, for all $u_1, u_2 \in U^i_{s-1}$, $|s_{l}(u_1) - s_{l}(u_2)| \leq 1$. We show that $u_1, u_2 \in U^i_s$, $|s_{l+1}(u_1) - s_{l+1}(u_2)| \leq 1$.

There are several possible cases:
\begin{enumerate}
    \item $u_1, u_2 \in U^i_{s-1}$ and $\sigma_{l+1}(u_1) = \sigma_{l+1}(u_2)$. Following the induction assumption, $|s_{l+1}(u_1) - s_{l+1}(u_2)| \leq 1$.
    \item $u_1, u_2 \in U^i_{s-1}$ but $\sigma_{l+1}(u_1) \neq \sigma_{l+1}(u_2)$ and $s_l(u_2) = s_l(u_1)$. By the definition of $x$-approval, $|s_{l+1}(u_1) - s_{l+1}(u_2)| = 1$. 
    \item $u_1, u_2 \in U^i_{s-1}$ but $\sigma_{l+1}(u_1) \neq \sigma_{l+1}(u_2)$ and, with out loss of generality, $s_l(u_2) > s_l(u_1)$. By the algorithm construction, $\sigma_{l+1}(u_2) = 0$ and $\sigma_{l+1}(u_1) = 1$ and thus by the induction assumption, $s_{l+1}(u_2) = s_{l+1}(u_1)$.
    
    \item With out loss of generality, $u_1 \in U^i_{s-1}$ and $u_2 \notin U^i_{s-1}$. In addition, $\sigma_{l+1}(u_1) = \sigma_{l+1}(u_2)$. By the definition of $U^i_{s-1}$, 
    $s_{l}(u_2) \geq s_{l}(u_1)$, and since $\sigma_{l+1}(u_1) = \sigma_{l+1}(u_2)$ then $s_{l+1}(u_2) \geq s_{l+1}(u_1)$.
    Since $u_2 \in U^i_s$, then $\exists u \in U^i_{s-1}$ such the $s_{l+1}(u) \geq s_{l+1}(u_2)$. According to cases $1,2$ or $3$, $s_{l+1}(u) - s_{l+1}(u_1) \leq 1$. Combining the inequalities we get that $s_{l+1}(u_2) - s_{l+1}(u_1) \leq 1$.
    
    \item With out loss of generality, $u_1 \in U^i_{s-1}$ and $u_2 \notin U^i_{s-1}$. In addition, $\sigma_{l+1}(u_1) \neq \sigma_{l+1}(u_2)$. By the definition of $U^i_{s-1}$, 
    $s_{l}(u_2) \geq s_{l}(u_1)$. In addition, by the algorithm construction $\sigma_{l+1}(u_1) = 1$ and $\sigma_{l+1}(u_2) = 0$.
    There are two possible cases:
    \begin{itemize}
        \item $s_{l+1}(u_2) < s_{l+1}(u_1)$. That is, $s_{l}(u_2) = s_{l}(u_1)$ and thus $s_{l+1}(u_1) - s_{l+1}(u_2) = 1$
        \item $s_{l+1}(u_2) \geq s_{l+1}(u_1)$. Since $u_2 \in U^i_s$, then $\exists u \in U^i_{s-1}$ such that $s_{l+1}(u) \geq s_{l+1}(u_2)$. According to cases $1,2$ or $3$, $s_{l+1}(u) - s_{l+1}(u_1) \leq 1$. Combining the inequalities we get that $s_{l+1}(u_2) - s_{l+1}(u_1) \leq 1$.
    \end{itemize}

    \item $u_1, u_2 \notin U^i_{s-1}$.  
    Since $u_1,u_2 \in U^i_s$ then $\exists u_1',u_2' \in U^i_{s-1}$ such the $s_{l+1}(u_1') \geq s_{l+1}(u_1)$ and $s_{l+1}(u_2') \geq s_{l+1}(u_2)$.
    According to cases $4$ or $5$, 
    $s_{l+1}(u_1') - s_{l+1}(u_1) \leq 1$ and $s_{l+1}(u_2') - s_{l+1}(u_2) \leq 1$.
    Let $u=\argmax\{u_1',u_2'\}$. Then, $s_{l+1}(u) - s_{l+1}(u_1) \leq 1$ and $s_{l+1}(u) - s_{l+1}(u_2) \leq 1$, and thus $|s_{l+1}(u_2) - s_{l+1}(u_1)| \leq 1$. 
\end{enumerate}

The proof for the set $D^i$ is similar.
\qed \end{proof}

\begin{lemma} \label{lemma:k_lexi_difference_of_candidates}
Given $i$, $1 \leq i \leq \lceil m/2 \rceil$, for all $u_1, u_2 \in U^i$, if $s_k(u_1) = s_k(u_2) + 1$ then $u_2$ is preferred over $u_1$ according to the lexicographical tie-breaking rule. Similarly, for all $d_1, d_2 \in D^i$, if $s_k(d_1) = s_k(d_2) + 1$ then $d_2$ is preferred over $d_1$ according to the lexicographical tie-breaking rule.
\end{lemma}

\begin{proof}
Recall that $U^i_0 = \argmin_{h \in H^i} \pos{h}{p_t}$, and for each $s=1,2,...$, $U^i_s=U^i_{s-1} \cup \set{u: u$ was ranked above some $u' \in U^i_{s-1}$ in some stage $l$, $1 \leq l < k$, but $u$ was ranked below some $u' \in U^i_{s-1}$ in stage $l + 1}$. In addition,  $U^i= \bigcup _{0 \leq s}U^i_s$.
We prove by induction on the index $s$, that for all $u_1, u_2 \in U^i$, if $s_k(u_1) = s_k(u_2) + 1$ then $u_2$ is preferred over $u_1$ according to the lexicographical tie-breaking rule. In the base case, when $s=0$, $|U^i_0|=1$ and thus the claim trivially holds. Assume that for $s-1$, for all $u_1, u_2 \in U^i_{s-1}$, if $s_{l}(u_1) = s_{l}(u_2) + 1$ then $u_2$ is preferred over $u_1$ according to the lexicographical tie-breaking rule. We show that $\forall u_1, u_2 \in U^i_s$, if $s_{l+1}(u_1) = s_{l+1}(u_2) + 1$ then $u_2$ is preferred over $u_1$ according to the lexicographical tie-breaking rule.
There are several possible cases:
\begin{enumerate}
    \item $u_1, u_2 \in U^i_{s-1}$ and $\sigma_{l+1}(u_1) = \sigma_{l+1}(u_2)$. Following the induction assumption, since $s_{l}(u_1) = s_{l}(u_2) + 1$ then $u_2$ is preferred over $u_1$ according to the lexicographical tie-breaking rule. Now, $s_{l+1}(u_1) = s_{l+1}(u_2) + 1$ and $u_2$ is preferred over $u_1$ according to the lexicographical tie-breaking rule.
    
    \item $u_1, u_2 \in U^i_{s-1}$ and $\sigma_{l+1}(u_1) \neq \sigma_{l+1}(u_2)$. Since $s_{l+1}(u_1) = s_{l+1}(u_2) + 1$ and $\sigma_{l+1}(u_1) \neq \sigma_{l+1}(u_2)$ then $s_{l}(u_1) = s_{l}(u_2)$. Since $s_{l+1}(u_1) = s_{l+1}(u_2) + 1$ then $\sigma_{l+1}(u_1) = 1$ and $\sigma_{l+1}(u_2) = 0$. That is, $u_2$ is preferred over $u_1$ according to the lexicographical tie-breaking rule (by the algorithm construction).
    
    \item $u_1 \in U^i_{s-1}$, $u_2 \notin U^i_{s-1}$. That is, $u_1$ was ranked lower than $u_2$ in stage $l$. However, $s_{l+1}(u_1) = s_{l+1}(u_2) + 1$, and thus  $\sigma_{l+1}(u_2) = 0$ and $\sigma_{l+1}(u_1) = 1$, and so $s_{l}(u_1) = s_{l}(u_2)$. Since $u_1$ was ranked lower than $u_2$ in stage $l$ then it must be that $u_2$ is preferred over $u_1$ according to the lexicographical tie-breaking rule.
    
    \item $u_1 \notin U^i_{s-1}$, $u_2 \in U^i_{s-1}$. Since $u_1 \in U^i_s$, then $\exists u \in U^i_{s-1}$ such that $u$ is ranked higher than $u_1$ in stage $l + 1$.
    Thus, $\sigma_{l+1}(u_1) = 0$ and $\sigma_{l+1}(u) = 1$.
    Since $u, u_1, u_2 \in U^i_s$, $s_{l+1}(u_1) = s_{l+1}(u_2) + 1$ and $u$ is ranked higher than $u_1$ in stage $l + 1$, by Lemma~\ref{lemma:k_score_difference_of_candidates}, $s_{l+1}(u_1) = s_{l+1}(u)$.
    Since $u$ is ranked higher than $u_1$ in stage $l + 1$, $u$ is preferred over $u_1$ according to the lexicographical tie-breaking rule.
    There are two possible cases: 
    \begin{itemize}
        \item $s_l(u) = s_l(u_2)$. That is, $\sigma_{l+1}(u_2) = 0$ and $\sigma_{l+1}(u) = 1$. Thus, $u_2$ is preferred over $u$ according to the lexicographical tie-breaking rule, by the algorithm construction. Therefore, $u_2$ is preferred over $u_1$ according to the lexicographical tie-breaking rule.
        
        \item $s_l(u) = s_l(u_2) + 1$. Following the induction assumption, since $s_l(u) = s_l(u_2) + 1$ then $u_2$ is preferred over $u$ according to the lexicographical tie-breaking rule. Therefore, $u_2$ is preferred over $u_1$ according to the lexicographical tie-breaking rule.
    \end{itemize}
    
    \item $u_1, u_2 \notin U^i_{s-1}$. Since $u_1, u_2 \in U^i_s$, by definition of $U^i_s$, $\exists u \in U^i_{s-1}$ such that $u$ is ranked higher than $u_1$ and $u_2$ in stage $l + 1$, but $u$ was ranked lower than $u_1$ and $u_2$ in stage $l$. Thus, $\sigma_{l+1}(u_1) = \sigma_{l+1}(u_2) = 0$ and $\sigma_{l+1}(u) = 1$. Since $s_{l+1}(u_1) = s_{l+1}(u_2) + 1$ and $\sigma_{l+1}(u_1) = \sigma_{l+1}(u_2)$, then $s_{l}(u_1) = s_{l}(u_2) + 1$. Therefore, $s_{l}(u) = s_{l}(u_2)$ and $s_{l+1}(u) = s_{l+1}(u_1)$. That is, $u$ is preferred over $u_1$ according to the lexicographical tie-breaking order. In addition, since $s_{l}(u) = s_{l}(u_2)$ and $u \in U^i_{s-1}$ and $u_2 \notin U^i_{s-1}$, then $u_2$ is preferred over $u$ according to the lexicographic tie-breaking order. Thus, $u_2$ is preferred over $u_1$ according to the lexicographical tie-breaking rule.
\end{enumerate}
The proof for the set $D^i$ is similar.
\qed \end{proof}

\begin{lemma} \label{lemma:U=1_or_D=1}
Given $i$, $1 \leq i \leq \lceil m/2 \rceil$, the set $|U^i| = 1$ or $|D^i| = 1$. 
\end{lemma}

\begin{proof}
Recall that $|H^i|=i$, $U^i_0 = \set{u_0}$ where $u_0 = \argmin_{h \in H^i} \pos{h}{p_t}$ and $D^i_0 = \set{d_0}$ where $d_0= \argmax_{d \in O \setminus H^i} \pos{d}{p_t}$. Since we use $x$-approval, if $x \geq i$ then by Lemma~\ref{lemma:k_ranking} all of the candidates of $H^i$ get a score of $1$ in each stage. Therefore, there is no candidate from $H^i$ that is ranked lower than $u_0$, and thus $U^i = U^i_0$. On the other hand, if $x < i$ then by Lemma~\ref{lemma:k_ranking} all of the candidates of $O \setminus H^i$ get a score of $0$ in each stage. Therefore, there is no candidate from $D^i$ that is ranked higher than $d_0$, and thus $D^i = D^i_0$. 
\qed \end{proof}

\begin{lemma} \label{lemma:k_lower}  
Given $i$, $1 \leq i \leq \lceil m/2 \rceil$, let $d* \in D^i$ such that $\forall d \in D^i$, $d \neq d*$ and $d* \succ_{p_t^a} d$. Similarly, let $u* \in U^i$ such that $\forall u \in U^i$, $u \neq u*$ and $u \succ_{p_t^a} u*$. If there exists $i$ such that $u* \succ_{p_t^a} d*$, and there are $k$ manipulators then there is a manipulation that makes $p$ the SPE result, and Algorithm~\ref{alg:coalitional_manip} will find it.
\end{lemma}

\begin{proof}
Assume that there exists $i$ such that $u* \succ_{p_t^a} d*$.
We first show that for all $o \in O \setminus (H^i \cup D^i)$ and $\forall d \in D^i$, it holds that $d \succ_{p_t^a} o$.
Assume by contradiction that there exists $o \in O \setminus (H^i \cup D^i)$ and $d \in D^i$, such that $o \succ_{p_t^a} d$. Then, by the definition of $D^i$, $o \in D^i$, which is a contradiction to the choice of $o$.
Let $h* \in H^i$ such that $\forall h \in H^i$, $h \neq h*$ and $h \succ_{p_t^a} h*$. Now, since $u* \succ_{p_t^a} d*$, then for all $o \in O \setminus H^i$, $o \prec_{p_t^a} u* = h*$ (otherwise, $h*$ would have been part of $U^i$). Therefore, $\A_{(\calF(\vec{p} \cup \vec{p}_M))}^i=H^i$, and by definition of $H^i$, $\A_{(p_o)}^i \cap H^i = \set{p}$. That is, the algorithm finds a manipulation that makes $p$ the SPE result.
\qed \end{proof}

\begin{lemma} \label{lemma:k_average}
Given $i$, $1 \leq i \leq \lceil m/2 \rceil$, let $q(U^i)$ and $q(D^i)$ be the average score of candidates in $U^i$ and $D^i$ after $k$ stages, respectively. That is, $q(U^i) = {\frac{1}{ |U^i|}}\sum_{u \in U^i}s_{k}(u)$, $q(D^i) = {\frac{1}{|D^i|}}\sum_{d \in D^i}s_{k}(d)$.
Let $d* \in D^i$ such that $\forall d \in D^i$, $d \neq d*$ and $d* \succ_{p_t^a} d$. Similarly, let $u* \in U^i$ such that $\forall u \in U^i$, $u \neq u*$ and $u \succ_{p_t^a} u*$. If for every $i$, $d* \succ_{p_t^a} u*$, and there are $k$ manipulators then there is no manipulation that makes $p$ the SPE result, and the algorithm will return false.
\end{lemma}

\begin{proof}
Assume that there is a successful manipulation $\vec{p}_{m}$ with $k$ manipulators.
Let $i$ be the corresponding iteration in which $RC$ returns $p$.
Let $p_t^m = \calF(\vec{p} \cup \vec{p}_m)$, and let $s^m_{k}(c)$ be the score of a candidate $c$ in $p_t^m$. Since Algorithm~\ref{alg:coalitional_manip} (as proved in Lemma~\ref{lemma:k_ranking}) places all the outcomes $u \in U^i$ at the $|U^i|$ highest positions and the outcomes $d \in D^i$ at the $|D^i|$ lowest positions, then:
{\small 
\begin{equation}\label{eq:qu_x-approval}
q(U^i) = \frac{1}{|U^i|}(\sum_{u \in U^i}s_0(u)+ k \cdot \min\set{x,|U^i|}) 
\geq 
\end{equation}
$$
\frac{1}{|U^i|}(\sum_{u \in U^i}s^m_{k}(u))=:q^m(U^i)
$$
\begin{equation}\label{eq:qd_x-approval}
q(D^i) = {\frac{1}{|D^i|}}(\sum_{d \in D^i}s_0(d)+ k \cdot (\max\set{|D^i|, m-x} - (m - x))) 
\leq
\end{equation}
$$
{\frac{1}{|D^i|}} \sum_{d \in D^i}s^m_{k}(d)=:q^m(D^i)
$$
}
 %
%
Since $d* \succ_{p_t^a} u*$, then ,by Lemma~\ref{lemma:k_score_difference_of_candidates}, $\lceil q(D^i) \rceil \geq \lfloor q(U^i) \rfloor$.
Combining the equations we get $\lceil q^m(D^i) \rceil \geq \lfloor q^m(U^i) \rfloor$. 
Since $q^m(D^i)$ and $q^m(U^i)$ are averages then there is at least one $u \in U^i$ and one $d \in D^i$, such that $s^m_{k}(d) = \lceil q^m(D^i) \rceil$ and $\lfloor q^m(U^i) \rfloor = s^m_{k}(u)$. Therefore, $s^m_{k}(d) \geq s^m_{k}(u)$.
If $s^m_{k}(d) > s^m_{k}(u)$, then $d \succ_{p_t^m} u$.
Otherwise, $s^m_{k}(d) = s^m_{k}(u)$, and we show that $\exists d' \in D^i$ such that $d' \succ_{p_t^m} u$.
According to Lemma \ref{lemma:U=1_or_D=1} either $|U^i|=1$ or $|D^i|=1$, so assume that $|U^i|=1$ and thus $u=u*$.
There are three possible cases:
\begin{enumerate}
    \item The algorithm assigns a score of $0$ to all of the candidates in $D^i$. In this case, according to Lemma \ref{lemma:conditions_for_D=1_U=1}, $|D^i|=1$. That is, $d=d*$. Since $s^m_{k}(d) = s^m_{k}(u)$ then $s_{k}(d) = s_{k}(u)$ (according to Equations \ref{eq:qu_x-approval},\ref{eq:qd_x-approval}). Since $d \succ_{p_t^a} u$, then $d \succ_{p_t^m} u$.
    \item The algorithm assigns a score of $0$ to some of the candidates in $D^i$, and $\forall d' \in D^i$ $s_k(d') \geq s_k(u)$. If $\exists d'' \in D^i$ such that $s_k(d'') > s_k(u)$ then $d'' \succ_{p_t^m} u$. Otherwise, $s_k(d)=s_k(d*)$. Since $s^m_{k}(d*) = s^m_{k}(u)$ then $s_{k}(d*) = s_{k}(u)$ (according to Equations \ref{eq:qu_x-approval},\ref{eq:qd_x-approval}). Since $d* \succ_{p_t^a} u$, then $d* \succ_{p_t^m} u$.
    \item The algorithm assigns a score of $0$ to some of the candidates in $D^i$, but $\exists d' \in D^i$ $s_k(d') < s_k(u)$. If $\exists d'' \in D^i$ such that $s_k^m(d'') > s_k^m(u)$, then $d'' \succ_{p_t^m} u$. Otherwise, let $y$ be the number of candidates from $D^i$ that have the score of $s_k(u)$ according to $p_t^a$. By Equations \ref{eq:qu_x-approval},\ref{eq:qd_x-approval}, there are at least $y$ candidates $d'' \in D^i$ such that $s_k^m(d'') = s_k^m(u)$. By Lemma~\ref{lemma:k_lexi_difference_of_candidates}, there is at least one candidate that, $d'' \in D^i$ such that $s_k^m(d'') = s_k^m(u)$ and $d'' \succ_{p_t^m} u$.
\end{enumerate}

Since $\min_{u \in U^i}s_{k}(u) = \min_{h \in H^i}s_{k}(h)$ and $|H^i|=i$ then there exists $d \in D^i$ such that $d \in \A_{(p_t^m)}^i$. 
Let $D^{i\prime} = \set{d \in O \setminus H^i : d \in \A_{(p_t^m)}^i}$ and let $H^{i\prime}= \set{h \in H^i : h \notin \A_{(p_t^m)}^i}$, where $D^{i\prime} = \set{d_1,\ldots,d_w}$ and $H^{i\prime}=\set{h_1,\ldots,h_w}$.
Now, we switch between the candidates from $D^{i\prime}$ and the candidates from $H^{i\prime}$ in $\vec{p}_m$. That is, given a preference order $p^a \in \vec{p}_m$ let $p^{a\prime} \leftarrow p^a$ and then for all $1 \leq j \leq w$, $h_j$ is placed in $p^{a\prime}$ in $\pos{d_j}{p^a}$ and $d_j$ is placed in $p^{a\prime}$ in $\pos{h_j}{p^a}$, for $d_j \in D^{i\prime}$ and $h_j \in H^{i\prime}$. Let $\vec{p}_{m'}$ be the manipulation where for each $p^a \in \vec{p}_m, p^{a\prime} \in \vec{p}_{m\prime}$, let $p_t^{m\prime} = \calF(\vec{p} \cup \vec{p}_{m\prime})$, and let $q^{m\prime}(U^i)$ and $q^{m\prime}(D^i)$ be the average scores of candidates from $U^i$ and $D^i$, respectively, in $p_t^{m\prime}$. Clearly, Equations~\ref{eq:qu_x-approval}-~\ref{eq:qd_x-approval} hold for $q^{m\prime}(U^i)$ and $q^{m\prime}(D^i)$ as well. That is, $\lceil q^{m\prime}(D^i) \rceil \geq \lfloor q^{m\prime}(U^i) \rfloor$ and so, $\exists d \in D^i$ and $u \in U^i$ such that $d \succ_{p_t^{m\prime}} u$.
On the other hand, 
for all $h \in H^i \setminus \set{p}$ and $d \in D^{i\prime}$, $\pos{h}{p_t} > \pos{d}{p_t}$, by the definition of $H^i$, and $p \in \A_{(p_t^{m\prime})}^i$. Therefore, $\A_{(p_t^{m\prime})}^i = H^i$, and since $U^i \subseteq H^i$ then $\lfloor q^{m\prime}(U^i) \rfloor \geq \lceil q^{m\prime}(D^i) \rceil$, and so, $\forall d \in D^i$ and $\forall u \in U^i$, $u \succ_{p_t^{m\prime}} d$, a contradiction. 
Therefore, there is no manipulation that makes $p$ the SPE result. 
The proof for the case where $|D^i|=1$ is similar.
Finally, if Algorithm~\ref{alg:coalitional_manip} returns $\vec{p}_{M}$ then it is a successful manipulation. Thus, if there is no manipulation the algorithm will return false.
\qed \end{proof}

\begin{proof} [Proof of Theorem~\ref{thm:coalitional_k_approval}]
Clearly, Algorithm~\ref{alg:coalitional_manip} runs in polynomial time.
According to Lemma \ref{lemma:k_lower}, if  there exists $i$ such that $u* \succ_{p_t^a} d*$, then there is a manipulation that makes $p$ the SPE result, and Algorithm~\ref{alg:coalitional_manip} will find it. On the other hand, according to Lemma \ref{lemma:k_average} if for every $i$, $d* \succ_{p_t^a} u*$ then there is no manipulation that makes $p$ the SPE result (with $k$ manipulators), and the algorithm will return false. Thus, Algorithm~\ref{alg:coalitional_manip} correctly decides the CC-MaNego problem with $x$-approval.
\qed \end{proof}
Unlike with the family of $x$-approval rules, CC-MaNego is computationally hard with Borda. The reduction is from the Permutation Sum problem (as defined by Davies et al.~\cite{davies2011complexity}) that is $NP$-complete~\cite{yu2004minimizing}.
\begin{definition}[Permutation Sum]\label{def:permutationSum} Given $n$ integers $X_1 \leq \ldots \leq X_n$ where $\sum_{i=1}^{n}X_i = n(n+1)$, do two permutations $\sigma$ and $\pi$ of $1$ to $n$ exist such that $\sigma(i)+\pi(i)=X_i$ for all $1 \leq i \leq n$?
\end{definition}
%
\begin{theorem} \label{thm:CCMaNego_reduction}
CC-MaNego is \text{NP}-Complete with Borda.
\end{theorem}
\begin{proof}
Clearly, the CC-MaNego problem is in $NP$. 
Now, let $\ptm = \calF(\vec{p} \cup \vec{p}_M)$.
Given an instance of the Permutation Sum problem we build an instance of the CC-MaNego problem as follows. There are $2n+4$ outcomes: $x_1,\ldots,x_n$, which correspond to the integers $X_1,\ldots,X_n$, $y_1,\ldots,y_{n+1}$ and three outcomes $a$, $b$ and $c$. By Lemma $1$ from~\cite{davies2011complexity}, we can construct an election in which the non-manipulators cast votes such that: 
\[
p_t = (y_1, \ldots, y_{n+1}, b, x_1, \ldots, x_n, a, c),
\]
 and the corresponding scores are: 
\[
(4n + 7 + C, \ldots, 4n + 7 + C, 4n + 6 + C ,
\]
\[
4n+6+C-X_1,
\ldots, 4n+6+C-X_n, C, z),
\] where $C$ is a constant and $z \leq C$.
The preference order of $p_o$ is as follows:
\[
p_o = (x_1,\ldots,x_n,b,a,c,y_1,\ldots,y_{n+1})
\] 
We show that two manipulators can make the outcome $a$ the SPE result iff the Permutation Sum problem has a solution.

($\Leftarrow$) Suppose we have two permutations $\sigma$ and $\pi$ of $1$ to $n$ such that $\sigma (i) + \pi (i) = X_i$. Let $\sigma^{-1}$ be the inverse function of $\sigma$, i.e., $i=\sigma^{-1}(x)$ if $x=\sigma(i)$. We define $\pi^{-1}(x)$  similarly.
We construct the following two manipulative votes:
\[
(a,y_1,\ldots,y_{n+1},c,x_{\sigma^{-1}(n)},\ldots,x_{\sigma^{-1}(1)},b)
\]
\[
(a,y_1,\ldots,y_{n+1},c,x_{\pi^{-1}(n)}, \ldots, x_{\pi^{-1}(1)},b)
\]
Since $\sigma (i) + \pi (i) = X_i$
and $z \leq C$, 
the preference profile $\ptm = \calF(\vec{p} \cup \vec{p}_M)$ is: 
\[
(y_1,y_2\ldots,y_{n+1},a,b,x_1, \ldots, x_n, c),
\]
since the corresponding scores are:
\[
(4n+7+C+2(2n+2),4n+7+C+2(2n+1),\ldots,
\]
\[
4n+7+C+2(n+2), 4n + 6 + C, 4n + 6 + C,
\]
\[
4n+6+C-X_1+X_1, \ldots, 4n+6+C-X_n+X_n, 2(n + 1) + z).
\]
Therefore, $\calN_t(\ptm,p_o) = \calN_o(\ptm,p_o) = a$.

($\Rightarrow$) Assume we have a successful manipulation. 
Such manipulation must ensure that all of the candidates $x_1,\ldots ,x_n$, and $b$ are not placed in the $n+2$ highest positions in $\ptm$, but $\pos{a}{\ptm} \geq n+2$.
That is, to ensure that outcome $a$ is ranked higher than outcome $b$, both manipulators have to place $a$ in the highest position in their preferences, and $b$ in the lowest position in their preferences. Thus, the score of outcome $a$ in $\ptm$ will be $4n+6+C$.
Let $\sigma(i)$ be a function that determines the score where the first manipulator assigned to outcome $x_i$. $\pi(i)$ is defined similarly for the second manipulator. 
Since the manipulation is successful, for every $i$, $1 \leq i \leq n$, 
$$
4n+6+C-X_i +\sigma (i)+ \pi (i) \leq 4n+6+C,
$$
and thus, 
$$
\sigma (i)+ \pi (i) \leq X_i.
$$
Since $\sum_{i=1}^n X_i = n(n+1)$,
$$
\sum_{i=1}^n \sigma(i)+\pi(i) \leq n(n+1).
$$
On the other hand, since $b$ is placed in the lowest position by both manipulators, 
$$
\sum_{i=1}^n \sigma(i) \geq \frac{n(n+1)}{2}
$$
and
$$
\sum_{i=1}^n \pi(i) \geq \frac{n(n+1)}{2}.
$$
Therefore, 
$\sum_{i=1}^n \sigma(i)+\pi(i) = n(n+1)$, and $\sum_{i=1}^n \sigma(i) = \sum_{i=1}^n \pi(i) = \frac{n(n+1)}{2}$. That is, $\sigma$ and $\pi$ are permutations of $1$ to $n$. 
Moreover, since there is no slack in the inequalities,
$$
\sigma (i)+ \pi (i) = X_i.
$$
That is, there is a solution to the Permutation Sum problem.
\qed
\end{proof}
Even though CC-MaNego with Borda is $NP$-complete, it 
might be still possible to develop 
an efficient heuristic algorithm that finds a  successful coalitional manipulation. We now show that Algorithm~\ref{alg:coalitional_manip} is such a heuristic, and show its theoretical guarantee. Specifically, the algorithm is guaranteed to find a coalitional manipulation in many instances, and we characterize the instances in which it may fail.
Formally,
\begin{theorem} \label{thm:mainTheorem}
Given an instance of CC-MaNego with Borda,
    \begin{enumerate}
        \item If there is no preference profile making $p$ the negotiation result, then Algorithm~\ref{alg:coalitional_manip} will return false. 
        \item If a preference profile making $p$ the negotiation result exists, then for the same instance with one additional manipulator, Algorithm~\ref{alg:coalitional_manip} will return a preference profile that makes $p$ the negotiation result.
    \end{enumerate}
\end{theorem}
That is, Algorithm~\ref{alg:coalitional_manip} will succeed on any given instance such that the same instance but with one less manipulator is manipulable. 
Thus, it can be viewed as a $1$-additive approximation algorithm (this approximate sense was introduced by \cite{zuckerman2009algorithms} when analyzing Borda as a social choice function (SCF)). 
%

Interestingly, the proof of Theorem~\ref{thm:mainTheorem} is in the same vein as the proof of Theorem~\ref{thm:coalitional_k_approval}, and we again use the sets $U^i$ and $D^i$. However, the proof here is more involved, and we begin by proving some Lemmas that are necessary for the proof.

\begin{lemma} \label{lemma:other_lower}
If there exists $i$ such that $\max_{d \in D^i}\set{s_{k}(d)} < \min_{u \in U^i}\set{s_{k}(u)}$, and there are $k$ manipulators then there is a manipulation that makes $p$ the SPE result, and Algorithm~\ref{alg:coalitional_manip} will find it.
\end{lemma}
\begin{proof}
Assume that these exists $i$ such that $\max_{d \in D^i}\set{s_{k}(d)} < \min_{u \in U^i}\set{s_{k}(u)}$.
We first show that for all $o \in O \setminus (H^i \cup D^i)$, it holds that $s_{k}(o) \leq min_{d \in D^i}\{s_{k}(d)\}$.
Assume by contradiction that there exists $o \in O \setminus (H^i \cup D^i)$ and $d \in D^i$, such that $s_{k}(o) > s_{k}(d)$. Then, by the definition of $D^i$, $o \in D^i$, which is a contradiction to the choice of $o$.
Now, since $\max_{d \in D^i}\set{s_{k}(d)} < \min_{u \in U^i}\set{s_{k}(u)}$, then for all $o \in O \setminus H^i$, $s_{k}(o) < \min_{u \in U^i}\set{s_{k}(u)} = \min_{h \in H^i}\set{s_{k}(h)}$ (otherwise, such $h \in H^i$ would have been part of $U^i$). Therefore, $\A_{(\calF(\vec{p} \cup \vec{p}_M))}^i=H^i$, and by definition of $H^i$, $\A_{(p_o)}^i \cap H^i = p$. That is, the algorithm finds a manipulation that makes $p$ the SPE result.
\qed \end{proof}

\begin{lemma} \label{lemma:average_min_max}
Given $i$, $1 \leq i \leq \lceil m/2 \rceil$, let $q(U^i)$ and $q(D^i)$ be the average score of candidates in $U^i$ and $D^i$ after $k-1$ stages, respectively. That is, $q(U^i) = {\frac{1}{ |U^i|}}\sum_{u \in U^i}s_{k-1}(u)$, $q(D^i) = {\frac{1}{|D^i|}}\sum_{d \in D^i}s_{k-1}(d)$.
If for every $i$, $q(U^i) < q(D^i)$, and there are $k-1$ manipulators then there is no manipulation that makes $p$ the SPE result, and the algorithm will return false.
\end{lemma}
\begin{proof}
Assume that there is a successful manipulation $\vec{p}_{m}$ with $k-1$ manipulators.
Let $i$ be the corresponding iteration in which $RC$ returns $p$.
Let $p_t^m = \calF(\vec{p} \cup \vec{p}_m)$, and let $s^m_{k-1}(c)$ be the Borda score of a candidate $c$ in $p_t^m$. Since Algorithm~\ref{alg:coalitional_manip} (as proved in Lemma~\ref{lemma:k_ranking}) places all the outcomes $u \in U^i$ at the $|U^i|$ highest positions and the outcomes $d \in D^i$ at the $|D^i|$ lowest positions, then:
{\small 
\begin{equation}\label{eq:qu}
q(U^i) = \frac{1}{|U^i|}(\sum_{u \in U^i}s_0(u)+ \sum_{j = 1}^{k-1} \sum_{i = 0}^{|U^i|-1} m-|U^i|+i) \geq 
\end{equation}
$$
\frac{1}{|U^i|}(\sum_{u \in U^i}s^m_{k-1}(u))=:q^m(U^i)
$$
\begin{equation}\label{eq:qd}
q(D^i) = {\frac{1}{|D^i|}}(\sum_{d \in D^i}s_0(d)+ \sum_{j = 1}^{k-1} \sum_{i = 0}^{|D^i|-1}i) \leq
\end{equation}
$$
{\frac{1}{|D^i|}} \sum_{d \in D^i}s^m_{k-1}(d)=:q^m(D^i)
$$
}
Combining the equations we get $q^m(D^i) > q^m(U^i)$. 
Since $q^m(D^i)$ and $q^m(U^i)$ are averages then there is at least one $u \in U^i$ and one $d \in D^i$, such that $s^m_{k-1}(d) \geq q^m(D^i)$ and $q^m(U^i) \geq s^m_{k-1}(u)$. Therefore, $s^m_{k-1}(d) > s^m_{k-1}(u)$. Since $\min_{u \in U^i}s_{k-1}(u) = \min_{h \in H^i}s_{k-1}(h)$ and $|H^i|=i$ then there exists $d \in D^i$ such that $d \in \A_{(p_t^m)}^i$. 
Let $D^{i\prime} = \set{d \in O \setminus H^i : d \in \A_{(p_t^m)}^i}$ and let $H^{i\prime}= \set{h \in H^i : h \notin \A_{(p_t^m)}^i}$, where $D^{i\prime} = \set{d_1,\ldots,d_w}$ and $H^{i\prime}=\set{h_1,\ldots,h_w}$.
Now, we switch between the candidates from $D^{i\prime}$ and the candidates from $H^{i\prime}$ in $\vec{p}_m$. That is, given a preference order $p^a \in \vec{p}_m$ let $p^{a\prime} \leftarrow p^a$ and then for all $1 \leq j \leq w$, $h_j$ is placed in $p^{a\prime}$ in $\pos{d_j}{p^a}$ and $d_j$ is placed in $p^{a\prime}$ in $\pos{h_j}{p^a}$, for $d_j \in D^{i\prime}$ and $h_j \in H^{i\prime}$. Let $\vec{p}_{m'}$ be the manipulation where for each $p^a \in \vec{p}_m, p^{a\prime} \in \vec{p}_{m\prime}$, let $p_t^{m\prime} = \calF(\vec{p} \cup \vec{p}_{m\prime})$, and let $q^{m\prime}(U^i)$ and $q^{m\prime}(D^i)$ be the average scores of candidates from $U^i$ and $D^i$, respectively, in $p_t^{m\prime}$. Clearly, Equations~\ref{eq:qu}-~\ref{eq:qd} hold for $q^{m\prime}(U^i)$ and $q^{m\prime}(D^i)$ as well. That is, $q^{m\prime}(D^i) > q^{m\prime}(U^i)$. On the other hand, 
for all $h \in H^i \setminus \set{p}$ and $d \in D^{i\prime}$, $\pos{h}{p_t} > \pos{d}{p_t}$, by the definition of $H^i$, and $p \in \A_{(p_t^{m\prime})}^i$. Therefore, $\A_{(p_t^{m\prime})}^i = H^i$, and since $U^i \subseteq H^i$ then $q^{m\prime}(U^i) \geq q^{m\prime}(D^i)$, a contradiction.
Therefore, there is no manipulation that makes $p$ the SPE result. Clearly, if Algorithm~\ref{alg:coalitional_manip} returns $\vec{p}_{M}$ then it is a successful manipulation. Thus, if there is no manipulation the algorithm will return false. 
\qed \end{proof}

\begin{definition}[due to \cite{zuckerman2009algorithms}]
\label{def:1_dense}
A finite non-empty set of integers $B$ is called $1$-dense if when sorting the set in a non-increasing order $b_1 \geq b_2 \geq \dots \geq b_i$ (such that $ \set{b_1, \dots, b_i}=B$),  $\forall j, 1 \leq j \leq i-1$, $b_{j+1} \geq b_j - 1$ holds.
\end{definition}
\begin{lemma} \label{lemma:1_dense}
Given $i$, $1 \leq i \leq \lceil m/2 \rceil$, let $U^i, D^i$ be as before. Then the sets $\set{s_{k-1}(u):u \in U^i}$ and $\set{s_{k-1}(d):d \in D^i}$ are $1$-dense.
\end{lemma}
\begin{proof}
Zuckerman et al.~\cite{zuckerman2009algorithms} 
define a set of candidates $G_W$, and show that the scores of the  candidates in $G_W$ are $1$-dense (Lemma $3.12$ in \cite{zuckerman2009algorithms}). Even though our definition of the set $D^i$ is slightly different, the set of scores $\set{s_{k-1}(d):d \in D^i}$ is essentially identical to the set of scores  of the candidates in $G_W$. Thus, it is $1$-dense. 
The proof for the set $\set{s_{k-1}(u):u \in U^i}$ is similar.
\qed \end{proof}

\begin{lemma} \label{lemma:average_fromMax_toMin}
Given $i$, $1 \leq i \leq \lceil m/2 \rceil$, $q(U^i) \leq \min_{u \in U^i}\set{s_{k}(u)} -m + |U^i|$, and similarly, $\max_{d \in D^i}\set{s_{k}(d)} \leq q(D^i) + |D^i| - 1$.
\end{lemma}
\begin{proof}
Sort the members of $U^i$ by their scores after stage $k-1$ in a decreasing order, i.e., $U^i = \set{u_1, \ldots , u_{|U^i|}}$ such that for all $1 \leq t \leq |U^i| - 1$, $s_{k-1}(u_{t+1}) \leq s_{k-1}(u_{t})$. Thus, $u_1 = \argmax_{u \in U^i}\set{s_{k-1}(u)}$.
Denote for $1 \leq t \leq |U^i|$, $g_t = s_{k-1}(u_t) + m - |U^i| + t - 1$, and let $G = \set{g_1, \ldots , g_{|U^i|}}$. Note that according to Algorithm~\ref{alg:coalitional_manip},  for any $t$, $s_{k}(u_t) = g_t$. Now, since the set $\set{s_{k-1}(u):u \in U^i}$ is $1$-dense (according to Lemma~\ref{lemma:1_dense}) then for any $1 \leq t \leq |U^i| -1$, $g_t \leq g_{t+1}$. Thus, for any $t >1$, $g_1 \leq g_t$. That is, $g_1 = \min_{u \in U^i}\set{s_{k}(u)} = s_{k-1}(u_1) + m - |U^i| = \max_{u \in U^i}\set{s_{k-1}(u)} + m - |U^i|$. Clearly, $q(U^i) \leq \max_{u \in U^i}\set{s_{k-1}(u)}$ since $q(U^i)$ is an average score, which is always less than or equal to the maximum score, and thus $q(U^i) \leq \min_{u \in U^i}\set{s_{k}(u)} -m + |U^i|$.

Similarly, sort the members of $D^i$ by their scores after stage $k-1$ in an increasing order, i.e., $D^i = \set{d_1, \ldots , d_{|D^i|}}$ such that for all $1 \leq t \leq |D^i| - 1$, $s_{k-1}(d_t) \leq s_{k-1}(d_{t + 1})$. Thus, $d_1 = \argmin_{d \in D^i}\set{s_{k-1}(d)}$. Denote for $1 \leq t \leq |D^i|$, $g_t = s_{k-1}(d_t) + |D^i| - t$, and let $G = \set{g_1, \ldots , g_{|D^i|}}$. Note that according to Algorithm~\ref{alg:coalitional_manip}, for any $t$, $s_k(d_t) = g_t$. Now, since the set $\set{s_{k-1}(d):d \in D^i}$ is $1$-dense (according to Lemma~\ref{lemma:1_dense}) then for any $1 \leq t \leq |D^i| - 1$, $g_t \geq g_{t+1}$. Thus, for any $t > 1$, $g_1 \geq g_t$. That is, $g_1 = \max_{d \in D^i}\set{s_k(d)} = s_{k-1}(d_1) +|D^i| - 1 = \min_{d \in D^i}\set{s_{k-1}(d)} + |D^i| - 1$. Clearly, $q(D^i) \geq \min_{d \in D^i}\set{s_{k-1}(d)}$, and thus $\max_{d \in D^i}\set{s_k(d)} \leq q(D^i) + |D^i| - 1$.
\qed \end{proof}
%
Now we can prove the theorem.
\begin{proof} [Proof of Theorem~\ref{thm:mainTheorem}]
Clearly, if Algorithm~\ref{alg:coalitional_manip} returns a preference profile $\vec{p}_M$, then it is a successful manipulation that will make $p$ the SPE result.
Suppose that a preference profile exists that makes $p$ the SPE result with $k-1$ manipulators. 
Then, by Lemma~\ref{lemma:average_min_max}, there exists $i$, $1 \leq i \leq \lceil m/2 \rceil$, such that 
$
q(D^i) + |D^i| - 1 \leq q(U^i) + |D^i| - 1.
$
By Lemma~\ref{lemma:average_fromMax_toMin},
$
q(U^i) + |D^i| - 1 \leq \min_{u \in U^i}\set{s_k(u)} -m +|U^i| + |D^i| - 1,
$
and 
$
\max_{d \in D^i}\set{s_{k}(d)} \leq q(D^i) + |D^i| - 1.
$
Since $|U^i| +|D^i| \leq m$,
$
\min_{u \in U^i}\set{s_k(u)} -m +|U^i| + |D^i| - 1 < \min_{u \in U^i}\set{s_k(u)}.
$
Overall, 
$
\max_{d \in D^i}\set{s_{k}(d)} < \min_{u \in U^i}\set{s_k(u)},
$
and by Lemma~\ref{lemma:other_lower} the algorithm will find a preference profile that will make $p$  the negotiation result with $k$ manipulators.
\qed \end{proof}

\section{Destructive Manipulation}
In this section we study the \textit{destructive} manipulation problem, where the goal of the manipulation is to prevent an outcome from being the SPE result.
We begin with the destructive variant of manipulation by a single voter.
\begin{definition}[D-MaNego]
    We are given a social welfare function $\calF$, a preference profile $\vec{p}$ of honest voters on the negotiating team $t$, the preference of the other party $p_o$, a specific manipulator $v'$, and a disliked candidate $e \in O$. We are asked whether a preference order $p_{v'}$ exists for the manipulator $v'$ such that $e \neq \calN_t(\calF(\vec{p} \cup p_{v'}),p_o)$ and $e \neq \calN_o(\calF(\vec{p} \cup p_{v'}),p_o)$.
\end{definition}

Recall that C-MaNego is in $P$ for any scoring rule, but this does not immediately imply that D-MaNego is also in $P$. Indeed, it is possible to run Algorithm~\ref{alg:single_manip} for each candidate $c \neq e$. However, since Algorithm~\ref{alg:single_manip} returns a manipulation only when $\calN_t(\calF(\vec{p} \cup p_{v'}),p_o) = \calN_o(\calF(\vec{p} \cup p_{v'}),p_o) = c$, it does not find a solution where $\calN_t(\calF(\vec{p} \cup p_{v'}),p_o) = c$ and $\calN_o(\calF(\vec{p} \cup p_{v'}),p_o) = c'$, $c \neq c'$, and both $c,c' \neq e$, which is a possible solution for D-MaNego. Nevertheless, we can use a slightly modified version of Algorithm~\ref{alg:single_manip} for D-MaNego.
\begin{theorem}
\label{thm:d_manego_p}
    D-MaNego with any positional scoring rule can be decided in polynomial time.
\end{theorem}
\begin{proof}
We use Algorithm~\ref{alg:single_manip} with the following changes. We change lines~\ref{line:1}-\ref{line:2} to check whether $e$ cannot be the SPE result. That is, if $\pos{e}{p_o} < \lfloor m/2 \rfloor$ the algorithm returns true, since every preference order $p_{v'}$ is a successful manipulation. In addition, we define $H^i$ as follows. Given $i$, $1 \leq i \leq \lceil m/2 \rceil$, let $p^* \neq e$ be the most preferred outcome in $p_t$ that belongs to $\A^i_{(p_o)}$. The set $H^i$ is composed of $p^*$ and the other $i-1$ most preferred outcomes in $p_t$ that are not $e$. This definition of $H^i$ is to ensure that there will be at least one outcome from $\A^i_{(p_o)}$ in the $i$ highest positions in $p_t^a$ while $e$ will not be in the $i$ highest positions in $p_t^a$. Finally, we place the remaining outcomes so that they will not make $e$ the negotiation result. Therefore, we change the condition in line~\ref{line:alg1_end_cond} to check whether $e \neq \calN_t(\calF(\vec{p}\cup p^a),p_o)$ and $e \neq \calN_o(\calF(\vec{p}\cup p^a),p_o)$. Following these changes the proof of correctness is similar to the proof of Theorem~\ref{thm:alg_1_correct}. In essence, whenever the proof of C-MaNego shows that $p \in \A^i_{(p_t^a)}$ we can show in the setting of D-MaNego that $e \notin \A^i_{(p_t^a)}$ but that there is another outcome $o \in \A^i_{(p_t^a)}$ and $o \in  \A^i_{(p_o)}$.
\qed
\end{proof}

We now continue with the destructive coalitional manipulation problem, where several manipulators might decide to collude and coordinate their votes in such a way that an agreed candidate $e$ will not be the SPE result. The problem is defined as follows:
\begin{definition}[DC-MaNego]
    Given a social welfare function $\calF$, a preference profile $\vec{p}$ of honest voters on the negotiating team $t$, the preference of the other party $p_o$, a number of manipulators $k$, and a disliked candidate $e \in O$,  we check whether a preference profile $\vec{p}_M$ exists for the manipulators such that $e \neq \calN_t(\calF(\vec{p} \cup \vec{p}_M),p_o)$ and $e \neq \calN_o(\calF(\vec{p} \cup \vec{p}_M),p_o)$.
\end{definition}

Similar to C-MaNego, we show that a slightly modified version of Algorithm~\ref{alg:coalitional_manip} decides DC-MaNego with any $x$-approval rule.
\begin{theorem}
\label{thm:d_manego_x-approval}
    DC-MaNego with any $x$-approval rule can be decided in polynomial time.
\end{theorem}
\begin{proof}
We use Algorithm~\ref{alg:coalitional_manip}, and change it in the same way that we change Algorithm~\ref{alg:single_manip} in the proof of Theorem~\ref{thm:d_manego_p}. Specifically, in lines~\ref{line:check_possibility}-~\ref{line:not_possible} we return true if $\pos{e}{p_o} < \lfloor m/2 \rfloor$, the set $H^i$ is composed of $p^*$ and the other $i-1$ most preferred outcomes in $p_t$ that are not $e$, and in line~\ref{line:check_success} we check if $e \neq \calN_t(\calF(\vec{p}\cup\vec{p}_M),p_o)$ and $e \neq \calN_o(\calF(\vec{p}\cup\vec{p}_M),p_o)$. Following these changes the proof of correctness is similar to the proof of Theorem~\ref{thm:coalitional_k_approval}. Specifically, Lemmas~\ref{lemma:k_ranking}, \ref{lemma:conditions_for_D=1_U=1}, \ref{lemma:k_score_difference_of_candidates}, \ref{lemma:k_lexi_difference_of_candidates} and \ref{lemma:U=1_or_D=1} still hold in the DC-MaNego setting. The proofs of Lemmas~\ref{lemma:k_lower} and \ref{lemma:k_average} are slightly changed, where instead of the claim that $\A_{(p_o)}^i \cap H^i = p$, we use the claim that $e \notin \A_{(p_o)}^i \cap H^i$ and $\A_{(p_o)}^i \cap H^i$ is not empty.
\qed
\end{proof}

Indeed, DC-MaNego with Borda is computationally hard.
Note that this result is surprising, since the destructive coalitional manipulation problem when using Borda as an SCF is in $P$~\cite{conitzer2007elections}. 
\begin{theorem} \label{thm:DCMaNego_reduction}
DC-MaNego with Borda is \text{NP}-Complete.
\end{theorem}
\begin{proof}
Clearly, the DC-MaNego problem is in $NP$. The proof of the $NP$-hardness is by a reduction from the Permutation Sum problem (definition~\ref{def:permutationSum}). 

Given an instance of the Permutation Sum problem we built an instance of the DC-MaNego problem as follows. There are $n + 4$ outcomes: $x_1,\ldots,x_n$, which correspond to the integers $X_1,\ldots,X_n$, $b_1,b_2$ and two outcomes $d$ and $e$. By Lemma $1$ from~\cite{davies2011complexity}, we can construct an election in which the non-manipulators cast votes such that: 
\[
p_t = (e, d, x_1, \ldots, x_n, b_1, b_2),
\]
 and the corresponding scores are: 
\[
(4n + 13 + C, 2n + 6 + C, 2n + 6 + C - X_1, \ldots, 2n + 6 + C - X_n, 
\]
\[
C, y), 
\]
where $C$ is a constant and $y < C$. The preference order of $p_o$ is as follows:
\[
p_o = (b_1, b_2, e, x_1,\ldots,x_n,d)
\] 
We show that two manipulators can prevent the outcome $e$ from being the SPE result iff the Permutation Sum problem has a solution.

($\Leftarrow$) Suppose we have two permutations $\sigma$ and $\pi$ of $1$ to $n$ such that $\sigma (i) + \pi (i) = X_i$. Let $\sigma^{-1}$ be the inverse function of $\sigma$, i.e., $i=\sigma^{-1}(x)$ if $x=\sigma(i)$. We define $\pi^{-1}(x)$  similarly.
We construct the following two manipulative votes:
\[
(b_1, b_2, e , x_{\sigma^{-1}(n)}, \ldots, x_{\sigma^{-1}(1)}, d)
\]
\[
(b_1, b_2, e, x_{\pi^{-1}(n)}, \ldots, x_{\pi^{-1}(1)}, d)
\]
Since $\sigma (i) + \pi (i) = X_i$
and $y < C$, 
the preference profile $\ptm = \calF(\vec{p} \cup \vec{p}_M)$ is: 
\[
(e, b_1, d, x_1, \ldots, x_n, b_2)
\]
since the corresponding scores are:
\[
(4n + 13 + C + 2(n+1), 2n + 6 + C, 2n + 6 + C, 2n + 6 + C, \ldots,
\]
\[
2n + 6 + C, y + 2(n + 2)).
\]
Therefore, $\calN_t(\ptm,p_o) = \calN_o(\ptm,p_o) = b_1 \neq e$.

($\Rightarrow$) Assume we have a successful manipulation. Clearly, every outcome $o \in \set{x_1, \ldots ,x_n,d}$ cannot be the SPE result since $e \succ_{p_o} o$. In addition, $b_2$ cannot be the SPE result, since in every possible manipulation $d \succ_{p_t^M} b_2$. Thus, outcome $b_1$ is the SPE result. Now, in every possible manipulation $e$ is placed in the highest position in $p_t^M$ due to its score in $p_t$. In addition, since $b_1$ is the SPE result it must be in the second highest position in $p_t^M$. Therefore, both manipulators have to place $b_1$ in the highest position in their preferences, and $d$ in the lowest position in their preferences.
Let $\sigma(i)$ be a function that determines the score that the first manipulator assigned to outcome $x_i$. $\pi(i)$ is defined similarly for the second manipulator. 
Since the manipulation is successful, 
$$
2n+ 6+C-X_i +\sigma (i)+ \pi (i) \leq 2n + 6 +C,
$$
and thus, 
$$
\sigma (i)+ \pi (i) \leq X_i.
$$
Since $\sum_{i=1}^n X_i = n(n+1)$,
$$
\sum_{i=1}^n \sigma(i)+\pi(i) \leq n(n+1).
$$
On the other hand, since $d$ is placed in the lowest position by both manipulators, 
$$
\sum_{i=1}^n \sigma(i) \geq \frac{n(n+1)}{2}
$$
and
$$
\sum_{i=1}^n \pi(i) \geq \frac{n(n+1)}{2}.
$$
Therefore, 
$\sum_{i=1}^n \sigma(i)+\pi(i) = n(n+1)$, and $\sum_{i=1}^n \sigma(i) = \sum_{i=1}^n \pi(i) = \frac{n(n+1)}{2}$. That is, $\sigma$ and $\pi$ are permutations of $1$ to $n$. 
Moreover, since there is no slack in the inequalities,
$$
\sigma (i)+ \pi (i) = X_i.
$$
Namely, there is a solution to the Permutation Sum problem.
\qed
\end{proof}
Finally, similar to CC-MaNego, we show that the modified  Algorithm~\ref{alg:coalitional_manip} is an efficient heuristic algorithm that finds a successful destructive manipulation, and we guarantee the same approximation. That is, the algorithm succeeds in finding a destructive manipulation for any given instance such that success for the same instance with one less manipulator is possible.
\begin{theorem} \label{thm:destructivemainTheorem}
There is a $1$-additive approximation algorithm for DC-MaNego with Borda.
\end{theorem}
\begin{proof}
We use Algorithm~\ref{alg:coalitional_manip}, and change it in the same way that we change Algorithm~\ref{alg:single_manip} in the proof of Theorem~\ref{thm:d_manego_p}. Specifically, in lines~\ref{line:check_possibility}-~\ref{line:not_possible} we return true if $\pos{e}{p_o} < \lfloor m/2 \rfloor$, the set $H^i$ is composed of $p^*$ and the other $i-1$ most preferred outcomes in $p_t$ that are not $e$, and in line~\ref{line:check_success} we check if $e \neq \calN_t(\calF(\vec{p}\cup\vec{p}_M),p_o)$ and $e \neq \calN_o(\calF(\vec{p}\cup\vec{p}_M),p_o)$. Following these changes the proof of correctness and the approximation is guaranteed similar to the proof of Theorem~\ref{thm:mainTheorem}. Specifically, Lemmas~\ref{lemma:k_ranking}, \ref{lemma:1_dense} and \ref{lemma:average_fromMax_toMin} still hold in the DC-MaNego setting. The proofs of Lemmas~\ref{lemma:other_lower} and \ref{lemma:average_min_max} are slightly changed, where instead of the claim that $\A_{(p_o)}^i \cap H^i = p$, we use the claim that $e \notin \A_{(p_o)}^i \cap H^i$ and $\A_{(p_o)}^i \cap H^i$ is not empty.
\qed
\end{proof}

\section{Conclusion and Future Work}
In this paper we analyze the problem of strategic voting in the context of negotiating teams. Specifically, a scoring rule is used as a SWF, which outputs an order over the candidates that is used as an input in the negotiation process with the VAOV protocol. 
We show that the single manipulation problem is in $P$ with this two stage procedure, and the coalitional manipulation is also in $P$ for any $x$-approval rule. The problem of coalitional manipulation becomes hard when using Borda, but we provide an algorithm that can be viewed as a $1$-additive approximation for this case.
Interestingly, our complexity results hold both for constructive and destructive manipulations, unlike the problems of manipulation when using Borda as an SCF.
Note also that our algorithms are quite general. Algorithm~\ref{alg:single_manip} provides a solution with any scoring rule. Algorithm~\ref{alg:coalitional_manip} solves the coalitional manipulation problem with any $x$-approval rule and it is also an efficient approximation with Borda.
%

For future work we would like to extend our analysis to other voting rules. In addition, designing FPT algorithms for CC-MaNego and DC-MaNego with Borda is a promising open research direction, since there is an FPT algorithm for the constructive coalitional manipulation of Borda as a SCF with respect to 
the number of candidates~\cite{yang2016exact}.

\section*{Acknowledgment}
This research was supported in part by the Ministry of Science, Technology \& Space, Israel.

\bibliographystyle{splncs04}
\bibliography{main}
\end{document}